\documentclass[conference]{IEEEtran}
\IEEEoverridecommandlockouts
\usepackage{url}

\usepackage{doi}

\usepackage{cite}

\usepackage{amsmath,amssymb,amsfonts}

\usepackage{algorithmic}

\usepackage[pdftex]{graphicx}
\DeclareGraphicsExtensions{.pdf,.jpeg,.png}
\graphicspath{{./.}}

\usepackage{textcomp}
\usepackage{xcolor}

\def\BibTeX{{\rm B\kern-.05em{\sc i\kern-.025em b}\kern-.08em
    T\kern-.1667em\lower.7ex\hbox{E}\kern-.125emX}}
\begin{document}

%%% S.M. S.N. S.S. O.P.N.

%%-------------------------------------------------------
\title{Attestation Infrastructures for Private Wallets}

\author{\IEEEauthorblockN{Thomas Hardjono}
\IEEEauthorblockA{MIT Connection Science \& Engineering \\
Cambridge, MA 02139, USA \\
hardjono@mit.edu}
}

\maketitle

%% Page numbering must be AFTER make-title
%
\pagestyle{plain} %--- Add page number to IEEE 2 column page (starts on second page)
\thispagestyle{plain}  %--- Add page number to first page

\begin{abstract}
In this paper we focus on one part of trust infrastructure
needed for the future virtual assets industry,
namely the {\em attestation infrastructure} related to key management
in private wallet systems.
Our focus is on {\em regulated private wallets} utilizing trusted hardware,
and the capability of the wallet to yield attestation evidence
suitable to address requirements in several use-cases,
such as asset insurance and regulatory compliance.
%%%
We argue that
attestation services will be needed as a core part of the key management lifecycle
for private wallets in true decentralized systems.
\end{abstract}


%%-------------------------------------------------

%%==========================================================
\section{Introduction}

The nascent virtual assets and cryptocurrencies industry
is today facing a number of challenges, 
not only related to their business models and to regulatory compliance,
but also challenges coming from the traditional banking and financial services sectors.
The entrance of banks into the virtual/digital assets space
will further shape the direction of the industry as whole.
%%%

Compared to the traditional financial industry -- which has developed and evolved
since the introduction of computerized trades in the 1960s --
the virtual assets industry still lacks the {\em trust infrastructures}
that permit decentralized systems such as blockchains and Digital Ledger Technology (DLT) systems
to compete with (and perhaps eventually replace) the legacy systems
in the traditional financial sector.

In this paper we focus on one part of trust infrastructures,
namely the {\em attestation infrastructure} related to key management
in private wallet systems (i.e. unhosted wallets).
Our focus is on the {\em regulated private wallet} utilizing trusted hardware,
and the capability for the wallet to attest evidence 
that can be used to address needs in several use-cases
and to satisfy regulatory compliance.
%%%
Thus, as part of the future trust infrastructure for decentralized systems,
attestation services will be needed as a core part of the key management lifecycle
for hardware wallets.
It is insufficient to merely deploy trusted hardware for wallets,
as there must be the means (mechanisms, protocols and standards) to: 
(i) permit wallets to truthfully report
its system state and the presence of certain types of keys inside the trusted hardware
(without disclosing the private keys)~\cite{TPM2003Design,TPM1.2-Arch},
and
for (ii) attestation verification services to correctly appraise these 
evidences~\cite{TCG-IWG-2005-Thomas-Ned-Editors-part1,TCG-IWG-2006-Thomas-Ned-Editors-part2}.

In the next section we provide some background
as to the state of centralized exchanges,
decentralized exchanges and the entrance of traditional banks into the 
virtual assets and crypto-currencies is market.
In Section~\ref{sec:Terminology} we provide
a short description of the terminology used in the paper.
%%%
We make the case for the attestations of private wallet systems
in Section~\ref{sec:UseCases}.

An attestation verification service offered by an entity
(e.g. VASP) should be part of a broader compliance management services
offered by the entity. 
This is discussed in Section~\ref{sec:VASPManagedCompliance}.
We explain the components or parts of the attestation infrastructure and services
in Section~\ref{subsec:InfrastructureAttestations},
followed by a high-level outline in Section~\ref{subsec:OverviewAttestationFlows}
of the process of attestations of a wallet.
In Section~\ref{subsec:VASP-Evidence-Types}
we describe some types of obtainable evidence that may be relevant
to third-parties, such as asset insurance providers.
We close the paper with some conclusions.

Our goal is to make this paper readable to a broad audience,
bearing in mind that device attestations based on trusted hardware
is a complex subject which is undergoing evolution.
Although new types of trusted hardware are entering the market,
we use the example of the humble TPM chip~\cite{TPM1.2specification}
in our discussion
because it incorporates many of the fundamental designs
in trusted computing~\cite{Proudler2002}, including attestations.
We assume the reader is at least familiar 
with how blockchain systems generally function.

%%==========================================================
\section{Background}
\label{sec:Background}

The Bitcoin model~\cite{Bitcoin} promised a peer-to-peer electronic cash system
that relied on a network of nodes,
essentially providing a distributed processing model.
The proposition is that instead of relying
almost exclusively on financial institutions -- serving as
trusted third parties to process electronic payments --
the use of the peer-to-peer payment system 
would offer decentralization of control away from these traditional financial institutions.
%%%

\subsection{CEX and Centralization}

The difficulty of managing keys on the part of end-users
has given rise to the {\em hosted wallet} (custodian) model,
where the asset service provider 
-- commonly referred to as {\em centralized crypto-exchanges} (CEX) --
holds and manages the private-public key pairs of their customers.
Current examples of centralized crypto-exchanges
include  Binance, Coinbase, Huobi Global and Kraken~\cite{CoinMarketCap2021Feb}.
From a regulatory perspective,
entities such as centralized crypto-exchanges
are referred to as 
{\em Virtual Asset Service Providers} (VASP)~\cite{FATF-Recommendation15-2018}.

The function of the CEX among others
is to hold and manage the user's private-public key pair
and utilize the key-pair on behalf of the user to sign transactions.
A further specialization is the case where 
the user (as the customer of the CEX) simply
opens an account at the CEX entity without being associated with
any user-specific private-public key pair.
In this {\em commingled accounts} arrangement, 
the CEX performs transactions on behalf of the customer
using the key-pair belonging to the CEX entity.

Needless to say,
the idea of a centralized CEX 
goes against the very heart of the decentralization of control
as envisioned by Nakamoto~\cite{Bitcoin}
and also by Chaum~\cite{Chaum81}.
%%%
The idea of ``peer-to-peer'' means that users must 
be able to deal with one another directly,
without the mediation of any third party.

\subsection{AML/CFT and Traditional Banks as New Entrants}

With the green light given by the US~OCC 
(Office of the Comptroller of the Currency) in July 2020
to allow national banks to provide 
cryptocurrency custody services for customers~\cite{OCC-Interpretive-Letter-1170},
many CEX entities now face the business challenge from  banks
entering the ``hosted wallet'' market~\cite{Lennon2020-ForbesOCC-crypto,De2020-OCC1170}.

The traditional banking and financial services sector has a number of advantages
over many newly-emerged CEX entities,
notably in the area of regulatory compliance in connection
with Anti-Money Laundering/Combating the Financing of Terrorism (AML/CFT).
%%%
One specific aspect of AML/CFT concerns to the need
for financial institutions to obtain, validate and exchange
customer information in the context of funds transfers and correspondent banking.
The Financial Action Task Force (FATF) -- as the
intergovernmental organization tasked with developing and evolving
policies to combat money laundering --
has placed {\em virtual assets}, including crypto-currencies,
under the same funds {\em Travel Rule} in its 2019 published
Recommendations~{No.~15}~\cite{FATF-Recommendation15-2018}
and guidelines~\cite{FATF-Guidance-2019}.
%%%

In practical terms,
this means that like traditional banks,
the CEX entities and other types of VASPs
must now obtain and validate their customer information
(for both legal persons and organizations)
and to share this information with each other
in the case of virtual asset transfers.
The customer sender and receiver information, among others, include:
(i) the name and account number of the originator (sender),
(ii) the address of the originator,
(iii) the amount and date of execution of the funds transfer,
(iv) the identity of the beneficiary's (recipient)
financial institution,
(v) the name and address of the beneficiary, 
(vi) the account number of the beneficiary and other beneficiary identifier,
and
(vii) the name, address and numerical identifier of the originator's 
financial institution~\cite{AlbertsNofziger2020-TravelRule}.
%%%
Compliance to the funds Travel Rule is already 
the expectation in the United States~\cite{TRISA2020-v07},
and the urgency of the matter is highlighted 
by the size and nature of activities related to cryptocurrencies 
(e.g. nearly three quarters of bitcoin that 
moved in exchange-to-exchange transactions 
was cross-border in the first half of 2020~\cite{CipherTrace-AML-Report-Spring2020}).

%%%
It is worth noting that traditional banks and financial institutions
have had to comply to the Travel Rule since at least the mid-1990s
due to the introduction of the Bank Secrecy Act (BSA) of 1996.
Thus,
over the past two decades
they have developed the information infrastructure 
to deal with the collection and verification of customer information.
They have also developed the infrastructure 
needed to securely deliver the originator and beneficiary
customer information in relation to international funds-transfers
and correspondent banking (e.g. SWIFT network~\cite{SwiftNet2004,SWIFT-CPS-2017}).
The hosted wallets model and the commingled-accounts model
can be readily replicated (and improved) by the traditional banks --
these institutions have been ``hosting'' people's money for centuries.
%%%

Thus, today it is insufficient for CEX entities and VASPs to merely be ``cryptocurrency transmitters''.
They must also develop new infrastructures
that assist them, among others, in complying to AML/CFT regulatory requirements.
At the same time,
these new infrastructures must also address the business challenge 
arising for the emerging 
{\em Decentralized Crypto Exchanges} (DEX)
-- which by definition require its participants
to hold their private-public key pair~\cite{Alkurd2020-DEX-Forbes}.

\subsection{Private Wallets: Asset Risk Management}

Fulfilling the vision of the decentralized 
peer-to-peer electronic cash system
as envisioned by Nakamoto~\cite{Bitcoin} 
necessitates the decentralization of control of private keys.
That is, it necessitates end-users holding and controlling their private keys
in their wallets with trusted hardware.
Aside from the protection of private-keys in these tamper-detectable hardware
and the key management lifecycle of these keys,
there are a number business challenges arising from the fact 
that private-keys (bound to virtual assets) are now distributed across 
hundreds of thousands (even millions) of endpoints.
One notable business challenge pertains to the ability
for these virtual/digital assets to be insured against theft and loss,
which in this case means risk of theft/loss of keys from millions of private wallets.

One promising avenue to begin addressing this asset insurance challenge
for private wallets is that of {\em device attestations} capabilities
offered by some types of trusted hardware 
(e.g. TPM~\cite{TPM1.2specification,TPM2.0specification,Proudler2014},
SGX~\cite{McKeen2016,McKeen2013}, 
TrustZone~\cite{ARM-TrustZone2009}, etc.).
Attestations is the process by which the trusted hardware (e.g. chip)
provides signed evidence regarding its internal state,
including the keys present in the shielded locations inside the hardware
without revealing the private-keys~\cite{CokerGuttman2011}.
This attestations capabilities permits external entities,
such as asset insurance providers,
to obtain assurance that (i) private-keys bound to virtual assets
are currently located in a given trusted hardware inside a wallet,
(ii) that the wallet is in possession of the user,
and (iii) that extracting the private-key from the trusted hardware
will be time consuming and economically costly for the attacker
who steals the wallet device.
%%%

These capabilities in-turn permit asset insurance providers
to more accurately perform risk assessment
over the virtual/digital assets tied to specific user wallets.
%%%
Factors that are of interest in risk assessment include,
but not limited to:
(a) the type of trusted hardware employed in the wallet,
(b) evidence that actual hardware is being utilized
(versus a virtualized~\cite{BergerCaceres2006-vTPM}
or software-emulated hardware~\cite{TPM-Automotive-Thin-Profile-2018}),
(c) the type of computer system within which the trusted hardware is utilized,
(d) known weaknesses and history of successful attacks on the family of trusted hardware,
and
(e) the current value of assets tied to the keys inside the trusted hardware.
Core to these functions is device attestations performed
by the trusted hardware as the attester.

Currently,
the market for hardware wallets is still nascent,
with a handful of products available for end-users
and enterprise customers
(e.g. Ledger's Nano~X~\cite{Ledger-Nano-X}
using Arm SecureCore~\cite{STMicro2020-SecurCoreSC300},
Silo from Metaco~\cite{Metaco2019}).
Several research efforts are continuing
on employing secure enclaves technology 
for wallet purposes (e.g.~\cite{Matetic2019-BITE-Bitcoin-SGX}).

On a historical note,
it is worth noting that low-cost cryptographic hardware have been available
for consumer PC computers since the mid-2000s,
in the form of the Trusted Platform Module (TPM) version 1.2 chip~\cite{TPM1.2specification}.
A landmark event occurred in 2006 when the US~Army mandated all its PC purchases
to contain the TPM chip~\cite{Gerber2016-Army-TPMs}.
%%%
This spurred many PC computer OEMs (such as Dell and Hewlett-Packard)
who sell into the US federal market
to begin incorporating the TPM chip inside their mid-range to high-end laptops
and PC computers.
This demand from the OEMs in turn
provided TPM hardware vendors (such as Intel, STMicro and Infineon)
with a return on their investment in trusted computing technologies,
which they had made
since 1999 when the Trusted Computing Group (TCG) industry consortium was established.
Applications of the TPM chip in PC computers
include file encryption (e.g. Microsoft BitLocker~\cite{BitlockerTPM}),
pairing with external encrypted disk-drives 
(e.g. Seagate Black Armor~\cite{TCG-OPAL-2009,Brandt2009-BlackArmorFDE}),
and protecting email signature keys~\cite{HardjonoKazmierczak2008}.
%%%
The current version is TPM version~2.0~\cite{TPM2.0specification,Proudler2014}.
%%%

Although more advanced trusted hardware technologies exist today
(e.g. Intel SGX~\cite{McKeen2016,McKeen2013}, 
Arm TrustZone~\cite{ARM-TrustZone2009}, 
Microsoft Pluton~\cite{Weston2020-MSFT-Pluton}),
the TPM chip is already widely available (and currently underutilized)
in several hundred million PC computers and related devices.

%----------------------fig--------------------------------------------
\begin{figure}[t]
\centering
\includegraphics[width=3.25in]{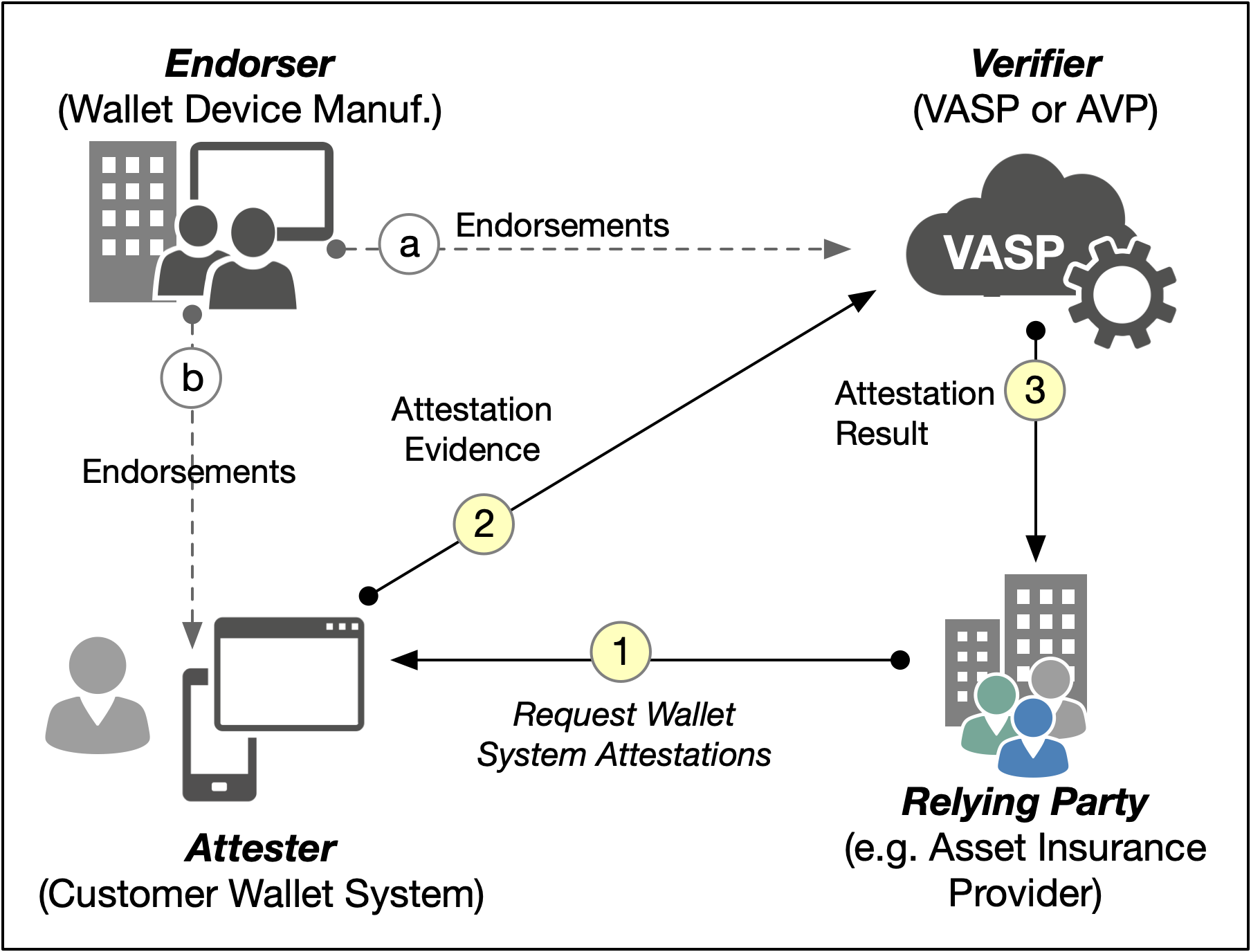}

\caption{Overview of Wallet Attestation Flows}
\label{fig:WalletAttestationFlows}
\end{figure}
%----------------------fig--------------------------------------------

%%==========================================================
\section{Terminology}
\label{sec:Terminology}

In the current work we use existing terminology as far as possible.
For readers seeking a basic definition of many of the terms 
related to blockchain technology,
we recommend the NIST guidance document on blockchain technology~\cite{NIST-8202-2018}.
\begin{itemize}

\item	{\em Virtual Asset Service Provider}: 
We use the definition of the VASP provided by FATF\cite{FATF-Recommendation15-2018}.
Virtual asset service provider means 
any natural or legal person who 
as a business conducts one or more of the following activities or 
operations for or on behalf of another natural or legal person:
(i) exchange between virtual assets and fiat currencies; 
(ii) exchange between one or more forms of virtual assets;
(iii) transfer of virtual assets;
(iv) safekeeping and/or administration of virtual assets or instruments enabling control over virtual assets; and
(v) participation in and provision of financial services related to an issuer's offer and/or sale of a virtual asset.
In this context of virtual assets, transfer means to conduct a transaction 
on behalf of another natural or legal person that moves 
a virtual asset from one virtual asset address or account to another.

\item	{\em Hosted wallets}: 
``A hosted wallet is an account-based software program for storing 
cryptographic keys controlled by an identifiable third party.
These parties receive, store, and transmit cryptocurrency transactions 
on behalf of their account holders; the
account holder generally does not have access to 
the cryptographic keys themselves''
(US~OCC Interpretive Letter~\#1172)~\cite{OCC-Interpretive-Letter-1172}.

\item	{\em Private wallets} (unhosted wallets): 
Broadly speaking, an {\em unhosted} wallet is 
one where the owner of a cryptocurrency maintains control of the
cryptographic keys for accessing the underlying cryptocurrency~\cite{OCC-Interpretive-Letter-1172}.

\item	{\em Regulated private wallets}: A private wallet is considered to be
{\em regulated} when the ownership
of the wallet is clear and provable using suitable technical means~\cite{FINMA-Guidance-2019}.
A regulated private wallet can be owned by an individual or organization
(e.g. corporation).

\item	{\em VASP-associated regulated private wallets}: 
A regulated private wallet is {\em associated} with a VASP
when the legal owner of the wallet has registered the wallet and public-key(s)
to an account at a regulated VASP.
The VASP must have obtained and validated the customer data for the account
(per the Travel Rule) before permitting the association
between the wallet and the account.
The term {\em regulated VASP} is used in the sense of FINMA~\cite{FINMA-Guidance-2019}.

\item	{\em Unverifiable private wallets}: A wallet whose legal ownership,
key-control and technological embodiment is unable to be verified.

\item	{\em Attestation}:
Attestation is a process for vouching for the accuracy of information 
that describes the properties and behavior of the target 
protected capabilities or shielded locations~\cite{TPM1.2specification,TCG-Glossary-2017}.  
The integrity of the target is established by checking 
the provenance of its origin and/or by verifying 
the veracity of state transitions from its original (origin) 
state to its current state~\cite{TCG-Attestations-Arch2020-Nov}. 
Attestations of a remote device often provides cryptographic evidence of the 
integrity of firmware, software and configuration that is operational in a device. 

\item	{\em Attestation Evidence}:
Evidence is typically in the form of an authenticated list of digests of values and actual values.  
Evidence must be unambiguously associated with the target device; 
this is often achieved using a cryptographic identifier that is physically bound to the device. 
Conceptually, attestation evidence is usually organized 
to provide proof of a series of secure operational 
state transitions from one trustworthy environment 
to another starting with the root of trust for 
measurement~\cite{TPM1.2specification}.

\item	{\em Attestation Verification}:
Verification of attestation evidence is determined 
by examining endorsements 
(i.e. manufacturer's cryptographic statements about the device,
such as its identity), 
then comparing signed evidence with authenticated expected values, 
all done in a separate trusted environment 
(referred to as the ``Verifier'')~\cite{TCG-Attestations-Arch2020-Nov}.

\end{itemize}

%%%%%%%%%%%%%%%%%%%%%%%%%%%%%%%%%%%%%%%%%%%%%%%%%%%%%%%%%%%%%%%%%%%%%%%%%%%%%%%%%%%%%%%%%%%%%%%%%%%%%%
%%%%%%%%%%%%%%%%%%%%%%%%%%%%%%%%%%%%%%% SECTION %%%%%%%%%%%%%%%%%%%%%%%%%%%%%%%%%%%%%%%%%%%%%%%%%%%%%%
\section{Making the Case for Attestations of Wallets}
\label{sec:UseCases}

One of the major challenges for the digital asset
industry relates to addressing the various attacks aimed at the client endpoints
which hold private-keys.
This need for key protection is part of the larger need for 
a comprehensive key management lifecycle~\cite{NIST-800-57,NIST-800-152} for 
the different types of scenarios and use-cases
related to blockchains generally.
The challenge can be expressed in the following, albeit simplistic, terms:
certain legitimate third-parties need some degree of {\em visibility into the state} of
a given wallet system (hardware, firmware, software)~\cite{NIST-800-193},
but {\em without visibility to the private-keys}
located within the wallet system.
Furthermore,
this must be performed without affecting the privacy 
of the user/owner of the wallet.

The following are two use-cases that are emerging in the near horizon:
\begin{itemize}

\item	{\em Asset insurance providers}:
An interesting case pertains to funds insurers~\cite{John2018,KharifLouis2018,Allison2020-insurance}
who may wish to enter the virtual assets and cryptocurrencies market
in order to expand their business coverage.
We refer to these entities as Asset Insurance Providers (AIP) generically.
Their primary interest is to obtain some visibility 
into the state of a target wallet as part of managing risks
associated with providing insurance to the assets bound to the keys in the wallet.
Clearly,
these third-parties must not obtain access to (see) the private-key in
the target wallet.
However,
they need to obtain assurance that:
(i) the private-key is present in the wallet
and that
(ii) sufficient protection is being utilized in the wallet
to protect the private-key (e.g. wallet uses real trusted hardware, not emulated).

\item	{\em CBDC Distributors}:
A crucial use-case coming on the horizon pertains to 
{\em Central Bank Digital Currencies} (CBDC)~\cite{AuerCornelli2020-CBDC,LiptonTreccani2021-book}
and the various {\em Stablecoins} that may be
derived from fiat digital currencies~\cite{Lipton2020b,LiptonSardon2020}.
%%%
Assuming that large financial institutions (e.g. commercial banks)
hold and distribute CBDCs to entities downstream (e.g. smaller regional banks),
these CDBC distributors will need to manage the cryptographic keys 
in enterprise-grade wallet systems (e.g. with Hardware Security Modules).
In this case,
the wallet systems of a CDBC-distributor must be able to yield
evidence that it is in a healthy state~\cite{IETF-rats-network-device-attestation-05},
operating using all the components (hardware, software, firmware)
it is designed to operate with.

\end{itemize}
Note that other forms of trusted hardware
(e.g.~\cite{TCG-DICE-Implicit-2018,Palmer-OCP2020-Attestation,Weston2020-MSFT-Pluton})
may be used for enterprise-level private wallets
which are designed to be non-mobile and be integrated into
the corporate directory services.

Emerging VASPs maybe in a good position today
to provide attestation verification services
for a variety of use-cases.
This is because in many use-cases there is a requirement that the verifier entity
be a {\em neutral} entity (not a wallet manufacturer, not a hardware/software vendor).
The attestation related services
should be part of a broader comprehensive managed compliance services from VASPs.

%%%%%%%%%%%%%%%%%%%%%%%%%%%%%%%%%%%%%%%%%%%%%%%%%%%%%%%%%%%%%%%%%%%%%%%%%%%%%%%%%%%%%%%%%%%%%%%%%%%%%%
%%%%%%%%%%%%%%%%%%%%%%%%%%%%%%%%%%%%%%% SECTION %%%%%%%%%%%%%%%%%%%%%%%%%%%%%%%%%%%%%%%%%%%%%%%%%%%%%%
\section{Managed Compliance Services}
\label{sec:VASPManagedCompliance}

\begin{figure}[t]
\centering
\includegraphics[width=3.25in]{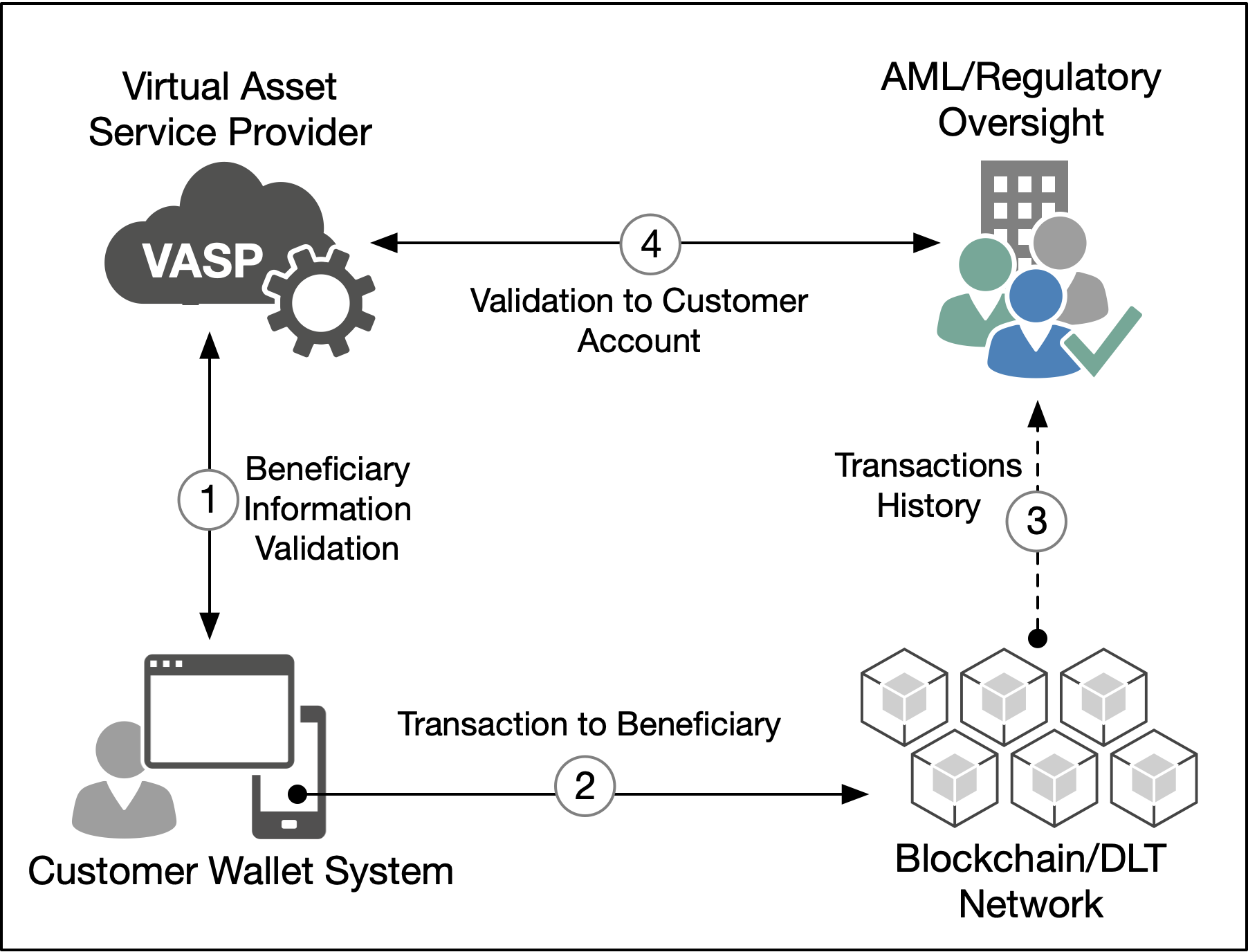}

\caption{Overview of VASP reconciliation of wallet-originated transactions}
\label{fig:WalletReconciliation}
\end{figure}

From the business opportunity perspective,
VASPs generally need to view their business function
as more than simply being cryptocurrency (asset) transmitters
-- namely the blockchain analog of the traditional wire-transfer shops.
VASPs need to identify new types of services
which can be offered to these private-wallet owners (both individuals and organizations)
that can provide business value to them.
There are several opportunities
which we place under the umbrella of {\em managed compliance} services.
Several types of offerings can be made
to these owners of private wallets:
\begin{itemize}

\item	{\em Certificate management services for public-keys}:
VASPs could offer digital certificate management services 
for the public-keys of the user employed on the blockchain
(i.e. to sign transactions destined for blockchain).
%%%
Acting in the role of a {\em Certification Authority}
the VASP can issue an {X.509} certificate~\cite{RFC2459-formatted,RFC3647-formatted}
for the transaction signing public-keys of the user.
Depending on the trusted hardware capabilities in the wallet,
the hardware can internally generate a private-public key pair,
followed by the user registering this public-key at the VASP
to obtain an {X.509} certificate.
%%%

~ ~This transaction public-key certificate
could be chained-up to the Root~CA certificate belonging
to the VASP itself (acting as the CA),
thereby permitting any recipient of the wallet-owner's {X.509} certificate
to identify the VASP that provided the certificate management services
to the owner of the private wallet~\cite{TRISA2020-v07}.

\item	{\em Travel Rule compliance services}:
One of the key challenges with regards to the Travel Rule
is the oft-occurring situation where the user (originator)
lacks information/data regarding the beneficiary.
For example,
the typical originator may only know the beneficiary's public-key,
name and email address 
(and perhaps their approximate geographic location).
Thus, for honest users employing private wallets,
the burden of complying to the Travel Rule lies with them -- something
that the ordinary citizen may find difficult and costly to achieve.
VASPs could play a crucial role in providing
Travel Rule related services to the owners of private wallets.

~ ~For example,
a VASP could offer services pertaining to the look-up
and validation of beneficiary information
in the context of transactions to be sent by the wallet.
%%%
For any given beneficiary address (public key),
the wallet can query the VASP's database prior
to transacting on the blockchain in order
to ensure that the beneficiary is a legitimate entity
(i.e. legal person or organization).

\item	{\em VASP tracking, audit and reconciliation services}:
For owners of private wallets who wish to comply
to the various regulations pertaining to virtual assets,
a VASP can perform tracking of all transactions
on the blockchain performed by one or more
keys in the private wallet.
This is notably relevant to private/permissioned 
blockchain and DLT systems.

~ ~Currently, many users employ token-tracking software
(e.g. EtherScan for the Ethereum platform)
which simply scans through the confirmed blocks of the ledger
to look for the user's public-key.
%%%
This manual process may not scale well
if the user is active on multiple blockchains simultaneously.
VASPs have the opportunity to assist these users
in performing audit and reconciliation
between the wallet key-usage log
with the history of user's transaction on these blockchains.
This audit process lends itself
to the VASP proving (e.g. to regulatory authorities) the compliant status
of the user (customer).
It also provides a means for the customer to satisfy taxation
requirements that may arise from asset transfers.
This is illustrated in Figure~\ref{fig:WalletReconciliation}.

\item	{\em Attestation services for private wallets}:
VASPs can provide attestation services 
-- notably the evidence verifier service --
to aid their customers in managing their private wallets.
This is discussed further in Section~\ref{subsec:OverviewAttestationFlows}.

\end{itemize}

%%%%%%%%%%%%%%%%%%%%%%%%%%%%%%%%%%%%%%%%%%%%%%%%%%%%%%%%%%%%%%%%%%%%%%%%%%%%%%%%%%%%%%%%%%%%%%%%%%%%%%
%%%%%%%%%%%%%%%%%%%%%%%%%%%%%%%%%%%%%%% SECTION %%%%%%%%%%%%%%%%%%%%%%%%%%%%%%%%%%%%%%%%%%%%%%%%%%%%%%

\section{Infrastructure for Attestations}
\label{subsec:InfrastructureAttestations}

Although the subject matter of attestations (or remote attestations)
is over two decades old now~\cite{TPM2003Design,Proudler2002},
it has only recently reached the mainstream narrative of system health in cybersecurity.
This interest has arisen on several fronts,
notably in the context of the {\em trustworthy supply chains} for the 
provenance of electronic components~\cite{Dodson2017,DodsonCabre2019,Benton2019}.
Given the spate of recent hacks (e.g.~\cite{NissenGronager2018,RobertsonRiley2018})
the provenance of components is core to cybersecurity more broadly.

In brief,
an attestation service requires at least the following components:
\begin{itemize}

\item	{\em Supply chain of endorsements}: Components manufacturers need to issue endorsements of their products
and to make these signed endorsements 
data-structures~\cite{TCG-IWG-2005-Thomas-Ned-Editors-part1,TCG-IWG-2006-Thomas-Ned-Editors-part2}
readily accessible to the verifier function
and the entity implementing verifier services.

\item	{\em Verifier function by attestation verification providers}: The function of the verification of evidences 
(yielded by a target attester device)
must be performed by a {\em neutral} entity, referred to here as the attestation verification provider (AVP).
The AVP must collect, collate and validate all the relevant endorsements
coming from the component manufacturers of systems.
Neutrality means that vendors cannot be an acceptable verifier for their own products.
Thus,
for example, the manufacturer of a PC computer (e.g. PC~OEM) or wallet system
should not be appraising its own endorsements.

~ ~Today the role of the attestation verification provider is still being defined,
and no such service exists as yet.

\item	{\em Attester capabilities in hardware components}: Certain types of components need to possess the attester capability
so that it can report on its integrity status.
Thus, for example,
future motherboards, system on chips (SoC) and related electronics should have native attester capabilities.
Currently,
efforts are underway in the semiconductor industry to begin addressing
the need for trusted supply chain traceability (e.g. Global Semiconductor Alliance (GSA-TIES)~\cite{GSA-TIES-2021}).

\end{itemize}

%%%%%%%%%%%%%%%%%%%%%%%%%%%%%%%%%%%%%%%%%%%%%%%%%%%%%%%%%%%%%%%%%%%%%%%%%%%%%%%%%%%%%%%%%%%%%%%%%%%%%%
%%%%%%%%%%%%%%%%%%%%%%%%%%%%%%%%%%%%%%% SECTION %%%%%%%%%%%%%%%%%%%%%%%%%%%%%%%%%%%%%%%%%%%%%%%%%%%%%%

\section{Summary of Attestations Process}
\label{subsec:OverviewAttestationFlows}

The need for better visibility into the state of a wallet system
is a common desire on the part of owners of private wallets
and of third-parties that provide insurance against
the loss and theft of virtual asset (i.e. loss/theft of keys).
Users as wallet owners want assurance that their wallet system
-- a complex computer system in its own right --
is behaving as designed and is performing its main function
(protecting keys).

Some legitimate and authorized third-parties
(such as asset insurance providers)
require in-depth proof regarding the status of private-keys in wallets,
beyond simply the classic proof-of-possession (PoP) of the private key --
which can easily be obtained
by running a challenge-response protocol
(e.g. CHAP protocol~\cite{rfc1994}) against the wallet.
Stronger proof is needed because the challenge-response protocol
only shows that the user is in possession of the private-key
-- it does not show {\em where} (i.e. which hardware) the private-key resides in.
A dishonest user can simply employ any PC computer
to sign the challenge (in the challenge-response protocol)
and then immediately claim loss of the private-key in order to repudiate
the signature.
Thus,
evidence is needed to show that the private-public key pair
(i) originated from within the trusted hardware,
(ii) that the private-key is not visible to the owner of the private-wallet
and (iii) that the private-key is not exportable from the trusted hardware
(i.e. the private-key is 
hardware-bound and non-migrateable~\cite{TPM1.2specification,HardjonoKazmierczak2008}).
Using the specific example of the TPM hardware,
some keys that are generated by the TPM can be configured upon creation
to be bound to that specific TPM instance.
This means that the TPM hardware will subsequently deny any request
to move or export the key from the TPM.
Such a key can also be used to ``certify'' application-level key pairs,
permitting these key pairs to be linked to the parent non-migrateable key.
A discussion on the TPM key hierarchy is beyond the scope
of the current work,
and the reader is directed to~\cite{HardjonoLipton2020-wallet-arxiv}
for further description in the context of wallets.

There are a number of technical-trust requirements that distinguish
systems which are {\em attestable} 
from those that are implicitly trusted.
The requirements include, among others,
that 
(1) the properties of the system that affect its behavior 
can be enumerated (i.e., trustworthiness attributes);
(2)	the trustworthiness attributes can be expressed in a machine-readable form;
(3) the vendors or other trusted sources can vouch for trustworthiness attributes;
and
that (4) the enumerated trustworthiness attributes can be reconciled with 
attributes provided by a trusted source~\cite{TCG-Attestations-Arch2020-Nov}.

An overview of the attestation flow of a wallet device is shown in 
Figure~\ref{fig:WalletAttestationFlows},
with the VASP taking the role of the {\em Verifier}
and an Asset Insurance Provider taking 
the role of the {\em Relying Party}~\cite{TCG-Attestations-Arch2020-Nov}.
The customer possessing the wallet is assumed
to have created an account at the VASP,
including registering the relevant public-keys (to be used by the wallet)
to that account.
In Step~1 the insurance provider requests that the user's
wallet system provide attestation {\em evidence}
regarding the wallet system state,
including keying material present in its trusted hardware.
This request could be regular (e.g. daily, weekly),
or it could be triggered by some action on the part of the user
(e.g. Insurance Provider notices large transaction on the blockchain
originating from the registered public-key).
In Step~2 the wallet system as the {\em attester}
yields a signed evidence and conveys it to the VASP as the verifier.
Note that a generic term for the verifier is also
the attestation verification provider (AVP).
In Step~3 the VASP as the verifier appraises the evidence from the attester,
and then delivers to the Relying Party
the {\em attestation result} based on the policies configured at the VASP.
An example of a generic attestation verification provider is outlined in~\cite{ZicHardjono2013}.

Also relevant in Figure~\ref{fig:WalletAttestationFlows}
is the signed {\em endorsements} from the wallet manufacturer 
(acting as the {\em endorser})
for the various components that constitute the wallet.
This is shown as Step~(a) and Step~(b),
and occurs out-of-band from the device attestation and evidence conveyance flows.

%%%%%%%%%%%%%%%%%%%%%%%%%%%%%%%%%%%%%%%%%%%%%%%%%%%%%%%%%%%%%%%%%%%%%%%%%%%%%%%%%%%%%%%%%%%%%%%%%%%%%%
%%%%%%%%%%%%%%%%%%%%%%%%%%%%%%%%%%%%%%% SECTION %%%%%%%%%%%%%%%%%%%%%%%%%%%%%%%%%%%%%%%%%%%%%%%%%%%%%%

%\section{Relevant Wallet Attestation Evidences for VASPs}

\section{Relevant Evidence from Wallets}
\label{subsec:VASP-Evidence-Types}

Depending on the specific type of trusted hardware,
there are a number of evidence types that can be
conveyed with the appropriate attestation protocol.
The following is a non-exhaustive list of some of the possible wallet and key information
that can be obtained using attestations~\cite{TCG-Attestations-Arch2020-Nov}:
\begin{itemize}

\item	{\em Key creation provenance}:
Most (if not all) current generation crypto-processor trusted hardware
have the capability to create/generate
a new private-public key pairs inside the shielded location
of the hardware, and to maintain keys inside its long-term non-volatile protected storage.
Furthermore, 
evidence regarding this process can be yielded by the trusted hardware,
allowing the provenance of such keys to be asserted.

~~Key-provenance evidence is useful for VASPs in many use-case scenarios.
For example,
in the case of a newly on-boarded customer bringing their existing wallet system, 
the VASP may wish to ascertain the provenance of the keys in that wallet.
If the provenance of the existing key-pair in the wallet is unverifiable,
then the VASP may require the customer (i.e. wallet)
to generate a new key-pair inside the wallet.
(assuming the wallet hardware is approved by the VASP).
This,
provides the VASP with a clear line of responsibility and accountability
(starting with the new key-pair in the customer wallet).
The VASP has exculpatory evidence regarding the on-boarding of the new customer
and the start of use of the new key-pair.

\item	{\em Evidence of geolocation of wallet}:
Depending on the type of trusted hardware used in the wallet, 
VASPs may be able to obtain evidence regarding the geolocation of a wallet device,
and therefore evidence regarding the geolocation of 
the hardware-bound keys in the wallet.
This may provide a means for VASPs to enforce geolocation-related
policies for customers to ensure that the VASPs customers
are operating within the permissible jurisdiction
(e.g. customer wallet must be in-country to sign transactions).
For example,
the work of~\cite{IETF-draft-ietf-rats-eat-03} includes the ability to
report location coordinates (latitude, longitude and altitude)
of the attester device.

\item	{\em Key usage sequence}:
VASPs can also make use of a number built-in features of some trusted hardware
via the application software (e.g. mobile app) on the wallet.
For example,
the application can use the underlying trusted hardware
to maintain a sequential history of the objects (transactions)
signed using the private key inside the trusted hardware
(e.g. in the TPM using the hash-extend operation with
the PCR registers and monotonic counter~\cite{TPM1.2specification,TPM2.0specification}).
This feature allows the VASP to perform an accurate accounting as to which
order transactions were signed by the wallet system,
as compared to the order in which the transactions were processed 
(i.e. confirmed) on the blockchain.

\item	{\em Evidence of wallet system configurations}:
Device attestations may permit a VASP to obtain
visibility into the device component compositions and configurations.
This may be useful information with regards to the 
{\em diversity}~\cite{NIST-8202-2018,HardjonoSmith2019c,Hardjono2021-GatewaysBridges} 
of wallet configurations,
something crucial from the perspective of malware 
aimed at wallet systems.
This has the advantage of allowing the VASP to advise (or require)
customers to replace a weak wallet system with a stronger system.

\end{itemize}

%----------------------fig--------------------------------------------
%----------------------fig--------------------------------------------

%%%%%%%%%%%%%%%%%%%%%%%%%%%%%%%%%%%%%%% SECTION %%%%%%%%%%%%%%%%%%%%%%%%%%%%%%%%%%%%%%%%%%%%%%%%%%%%%%

\section{Conclusions}

VASPs today face a number of challenges,
ranging from regulatory compliance related to virtual assets,
to potential competition arising from the
traditional banking sector.

The vision of Chaumian eCash and of Nakamoto's Bitcoin 
is that of a peer-to-peer electronic cash
based on decentralized control.
Contrary to the current dominance of the centralized exchange (CEX) model with hosted wallets,
this vision points to the need for end-users
to hold and control their private keys.
This, in turn, points to the requirement that the end-user's wallet system
incorporate trusted hardware
to protect the private keys.

However,
trusted hardware alone is not enough.
There must be an attestation infrastructure -- implementing 
standardized mechanisms and trust protocols -- that supports
the wallet in truthfully reporting its internal system state,
including the state of its trusted hardware and the presence of certain types of keys
in the trusted hardware.
This feature is relevant to legitimate third-party entities
-- such as asset insurance providers and auditors of CBDC wallets --
who require some degree of visibility into state of the wallet,
but without disclosure of the private keys in the wallet.

VASPs are best positioned today to provide neutral attestation verification services
for the different types of use-cases related to wallets.
%%%
These attestation related services
should be part of a more comprehensive managed compliance services
offered by VASPs more generally.

\section*{Acknowledgement}
We thank Sandy Pentland and Alex Lipton (MIT) for support
in this work.
We especially thank the various members of the Trusted Computing Group (TCG)
-- notably members of the Infrastructure Working Group
and the Attestations Working Group --
who over the past 20 years have worked tirelessly
to define and standardize trusted computing technologies.

%		\bibliographystyle{IEEEtran}
%		\bibliography{./biblio/IEEEabrv,./biblio/hardjonobib,./biblio/thomasrfcbib}


\end{document}